\documentclass[aps,prb,twocolumn,groupedaddress]{revtex4}
\usepackage{graphicx}
\begin{document}
\newcommand{\HH}{\mathbf{H}}
\newcommand{\DD}{\mathbf{D}}
\newcommand{\GG}{\mathbf{G}}
\newcommand{\ggamma}{\mathbf{\Gamma}}
\newcommand{\sss}{\mathbf{\Sigma}}
\newcommand{\PRB}{{\em Phys. Rev. B }}
\newcommand{\PRL}{{\em Phys. Rev. Lett. }}
\newcommand{\APL}{{\em App. Phys. Lett. }}
\newcommand{\SCI}{{\em science }}
\newcommand{\NAT}{{\em nature }(London) }
\newcommand{\JPCA}{{\em J. Phys. Chem. A}}
\newcommand{\NL}{{\em Nano Lett.}}
\newcommand{\CPL}{{\em Chem. Phys. Lett.}}
\newcommand{\JCP}{{\em J. Chem. Phys. }}
\newcommand{\dtb}{{$d_{TB} $}}
\newcommand{\dts}{{$d_{Tip-Sub} $}}
\newcommand{\A}{{\bf A }}
\newcommand{\B}{{\bf B }}
\newcommand{\dr}{${\bf D(r)}$}
\newcommand{\Pib}{$\pi$}
\newcommand{\EF}{$E_F $}
\newcommand{\EG}{$E_g $}
\author{H. Mehrez}
\author{A. Svizhenko}
\author{M. P.Anantram}
\affiliation{Mail Stop: 229-1, NASA Ames Research center, Moffett Field, CA 94035-1000, USA}
\author{M. Elstner}
\author{T. Frauenheim}
\affiliation{Theoretische Physik, Universitat Paderborn, D-33098 Paderborn, Germany} 
\date{\today}

\title{Analysis of band-gap formation in squashed
arm-chair CNT}

\begin{abstract}
The electronic properties of squashed arm-chair carbon nanotubes are
modeled using constraint free density functional tight binding
molecular dynamics simulations. Independent from CNT diameter,
squashing path can be divided into {\it three} regimes. 
In the first regime, the nanotube
deforms with negligible force. In the second one, there is
significantly more resistance to squashing with the force being 
$\sim 40-100$ nN/per CNT unit cell. In the last regime, the CNT looses
its hexagonal structure resulting in force drop-off followed by
substantial force enhancement upon squashing.
We compute the change in band-gap
as a function of squashing and our main results are: 
(i) A band-gap initially opens due to interaction between atoms at the
top and bottom sides of CNT.
The $\pi-$orbital approximation is successful in modeling the band-gap 
opening at this stage.
(ii) In the second regime of squashing, large $\pi-\sigma$ interaction 
at the edges becomes important, which can lead to band-gap oscillation. 
(iii) Contrary to a common perception, nanotubes with broken mirror 
symmetry can have {\it zero} band-gap. 
(iv) All armchair nanotubes become metallic in the third regime of squashing.
Finally, we discuss both differences and similarities obtained from
the tight binding and density functional approaches.  
\end{abstract}
\pacs{73.63.Fg,61.46.+w,73.22.-f}

\maketitle

\section{Introduction}

Experiments probing the electromechanical response of carbon 
nanotubes have been the most interesting recent work in
nanotubes.~\cite{paulson-APL-99,tombler,minot-prl-03,cao-prl-03,gomez2004}
These experiments involve nanotubes interacting electrically with 
contacts and mechanically with an atomic force microscope (AFM) tip. 
Apart from the fundamental physics governing the electromechanical 
response, these experiments are also important for future use of 
carbon nanotubes as actuators and nano electromechanical
devices.~\cite{Gartstein-PRL-02} 
There are two categories of experiments exploring the electromechanical properties.
The first category involves deformation of a suspended nanotube with
an AFM tip\cite{tombler,minot-prl-03,cao-prl-03}. The electrical
conductance was found to decrease by two orders of magnitude upon
modest deformation due to bond stretching which results in band-gap
opening at Fermi energy\cite{maiti2002}. The second category of
experiments involves the squashing of nanotubes lying on a hard
substrate\cite{paulson-APL-99,gomez2004}. 
In the experiment of Gomez {\it et al.}\cite{gomez2004} a metal to
semiconductor transition has been demonstrated in 
squashed metallic nanotube. Theoretically, there have been
several studies that modeled the electro-mechanical properties of
squashed carbon
nanotubes\cite{kilic2000,park1999,gulseren2002,lu2003}. It was
found that mirror symmetry breaking and formation of bonds between
atoms at the top-bottom sides of CNT are necessary to open a band-gap
in an arm-chair nanotube. While
reference [\onlinecite{park1999,gulseren2002}] performed
energy relaxation by enforcing specific symmetries during deformation,
reference [\onlinecite{lu2003}] modeled squashing by a tip whose
width was $5.8$ \AA, which is smaller than even a (6,6) nanotube
diameter. \\ 
In this study, we model the tip-nanotube interaction more
realistically by performing constraint free density functional tight
binding (DF-TB) molecular dynamics (MD) simulations. Our calculations differ
from prior work in that we do not impose specific symmetry conditions
and allow for the nanotube to change symmetry during
deformation. Further, our tip diameter is larger than the nanotube
diameter, as in the experiments of Ref.[\onlinecite{paulson-APL-99,gomez2004}]. The aim of
our work is to investigate:  
(i) the magnitude of the force required to squash CNT,
(ii) dependence of band-gap formation on the initial conditions of the
MD simulations,  
(iii) applicability of $\pi-$orbital theories to find the band-gap upon
deformation, 
(iv) relative roles of interactions between atoms at the top-bottom
sides and atoms at the edges on the band-gap,
(v) the electronic properties of a CNT squashed beyond the reversible
regime, 
(vi) diameter dependence on the band-gap opening upon squashing, and
(vii) effect of self consistent calculations on band-gap formation.\\
We will address the aforementioned issues in the rest of the
paper. Initially we describe our method to simulate the
experiment. Following this we present the mechanical properties of
squashed CNT in Section \ref{sec:mecpro}. In Section \ref{sec:elpro},
various aspects on the electronic properties of squashed CNT will be
investigated. Finally, 
the conclusion will be given in Section \ref{sec:conclu}.

\section{methodology  \label{sec:method}}
An $(n,n)$ CNT is generated from a planar rolled
graphene strip, where $n\in\{6,7,8,9,10,11,12\}$. The corresponding
diameters of these CNTs are in the range of $\sim 8-16$  \AA. If such tubes
are squashed in an AFM set-up, the tip radius is at least {\it one order}
of magnitude larger than the CNT diameter. Hence we approximate the tip as a
rigid graphene sheet where the atoms do not relax. The substrate is also 
modeled as a rigid graphene sheet. The distance between the graphene sheets
and CNT edges is at least $4.5$  \AA~ at the beginning of the MD
simulations, so that CNT is not deformed in the first step. 
A snap shot of the atomic configuration of a squashed CNT between
tip and sample is shown in the inset of Fig.\ref{distance-force}-b. To
model squashing of CNT we adopt two 
different methods: the first method consists of 
moving both the tip and substrate symmetrically towards the nanotube and
the second one consists of moving only the tip towards
the nanotube. We do not employ any further constraint on CNT atoms.\\
The MD simulations are performed using DFTB, which is a
density functional theory based tight binding
approach\cite{elstner1998,frauenheim2000}. At each MD step of
squashing, forces are calculated using
Hellmann-Feynman\cite{hellmann1937,feynman1938} theorem and CNT 
atoms are fully relaxed by the conjugate gradient method. Convergence
criteria on the relaxed CNT structure is set such that the  maximum
force of on each CNT atom is $<10^{-4}au$.  
We have squashed the nanotubes all the way to irreversible regime. 
That is,  the CNT
structure is lost at the end of the simulation and cannot be recovered
if the graphene sheets are removed. Since CNTs are constraint free,
both ``tank treading'' (motion of atoms along circumference) and
translational motion at various stages of squashing are possible if
the energetics is favorable to such movement. We have also found that 
the symmetry of squashed CNT depends on the initial conformation of the
CNT with respect to tip and substrate as well as the method adopted to
squash CNT.

As CNT is being squashed, atoms follow various
trajectories depending on the details of initial and ambient
conditions. This would affect the conformation of the CNT with respect
to the AFM tip as well as the symmetry of the CNT itself during
squashing.  
To model this experiment faithfully and to draw more general
conclusions on the electro-mechanical properties of squashed arm-chair
CNTs, various simulations are required. In our work, constraint free
squashing of more than {\it fifteen} paths have been performed on
various CNTs and this represents substantial improvement over previous
simulations\cite{park1999,gulseren2002,lu2003}.

\section{mechanical properties  \label{sec:mecpro}}
Our MD simulations reveal {\it three} different regimes
of a squashed nanotube. These regimes can be clearly resolved by plotting
the distance between top and bottom sides of the nanotube (\dtb) {\it
vs.} tip-substrate separation distance (\dts) as shown in
Fig.\ref{distance-force}-a as well as the force on the tip {\it vs.}
\dts~ as shown in Fig.\ref{distance-force}-b.  
In the first regime, \dtb~ is 
larger than $3.8$ \AA~ and the nanotube is compressed relatively
easily. This is reflected in the rather large slope
$\Delta$\dtb$/$$\Delta$\dts~$\sim 0.88 \pm 0.07$ as seen in
Fig.\ref{distance-force}-a, and small force of less than $10$ nN/per
CNT unit cell on the tip as seen in Fig.\ref{distance-force}-b.
The second regime of squashing corresponds to \dtb~$\sim 2.3-3.8$
 \AA. Here the nanotube is more rigid to compression as seen by the moderate value of $\Delta$\dtb$/$$\Delta$\dts $\sim 0.6\pm 0.05$, and the
relatively large value of force on the tip, $ 40-100$ nN/per 
CNT unit-cell. These two regimes are reversible, {\it i.e.} retracting the
tip and substrate results in a reversible change of the nanotube.
The third regime corresponds to \dtb~$<2.3$ \AA~ and is characterized by an 
irreversible change, where the honey-comb structure of the nanotube is
lost. The nanotube undergoes major atomic rearrangements so that total
energy of the system is reduced. The CNT structural relaxation is also
accompanied by force drop-off on the tip. However, this is
substantially enhanced upon further squashing as seen in
Fig.\ref{distance-force}-b. An interesting feature of this regime
is that atoms at the top and bottom sides of CNT are no longer planar. They
form a sawtooth structure and results in an increase of  
\dtb~ with decrease in \dts~as shown in Fig.\ref{distance-force}-a.

In Fig.\ref{distance-force}-b,c we present the force curves on the tip
for all CNTs as a function of \dts~and \dtb, respectively. The curves
corresponding to (6,6) and (12,12) nanotubes are shown as thick 
dotted and dashed lines, respectively. Comparison of the 
two curves indicates that, in general the force on the tip increases with
increase in nanotube diameter. However, we find that the force
required to squash an (11,11) nanotube (shown as thin lines) is larger than
the one needed for a (12,12) nanotube when \dtb~$>2.3$  \AA. The underlying
reason for this discrepancy is that the mirror symmetry is preserved
during this simulation of the (11,11) nanotube while broken in the
(12,12) CNT. If the nanotube mirror symmetry is preserved,
atoms at the top-bottom sides of the nanotube are aligned. This results in a
``more robust'' structure, which offers resistance to
squashing. Over all, our results indicate that the force required to squash
nanotubes increases with increase in diameter, and is larger for
conformation which preserves mirror symmetry.

\section{Electronic properties \label{sec:elpro}}

(6,6) arm-chair CNT has been investigated substantially and we start
our discussion on the electronic properties of this tube.
Depending on the initial orientation of 
the tube with respect to the 
graphene sheets and the manner of squashing, structural deformation may 
proceed along different paths.
At each step of these paths and when the CNT atoms
are fully relaxed, we compute the band structure and find the band-gap.
Fig.\ref{6-6-tube-gap} shows an evolution of the 
band-gap as a function of \dtb~ for different paths. 
Qualitatively, one can see common features corresponding to the three
regimes of squashing. 

In the first regime, when \dtb $> 3.8$  \AA, initially
metallic (6,6) CNT shows a zero band-gap. The transition 
to the second regime is manifested by the opening of the band-gap at
\dtb $\sim 3.8$  \AA, (Fig.\ref{6-6-tube-gap}-(b-d)).  
It is at this deformation  that atoms at the top and bottom sides of
the CNT start to interact, which leads to the stiffening of the tube 
and to the modification of the band structure.
The interaction between top and bottom sides of the CNT induces
a perturbation term to ideal CNT Hamiltonian $\HH_0$. We define this
additional term as $\Delta \HH_{TB}$ and it
corresponds to the interaction between atoms shown as grey and dark
grey atoms in the insets of each panel of Fig.\ref{6-6-tube-gap}. 
We note, that the opening of the band-gap may occur only 
when the mirror symmetry of the tube is broken, which is not the case in 
Fig.\ref{6-6-tube-gap}-a. 
 
In the second regime, where CNT is squashed such that $2.3$ \AA$<$ \dtb
$<3.8$ \AA, major changes to the band-gap occur. 
The most striking feature
is that while the band structure is qualitatively similar to that of the 
undeformed nanotube, {\it the band-gap oscillates}, 
becoming {\it zero}, as in Fig.\ref{6-6-tube-gap}-b,c at \dtb$\sim$2.90 \AA.
Another important observation is that the symmetry of 
the tube is constantly changing in the course of deformation. 
Sudden changes of symmetry are accompanied by steep changes of band-gap, as
in Fig.\ref{6-6-tube-gap}-d at \dtb$\sim$3.0 \AA.
The galore of various scenarios of band-gap behavior 
under squashing raises a 
question about the robustness and universality of 
the conductance modulation in such experiments.
There are two sources of uncertainty. 
The first one is the constantly changing symmetry 
of the tube which leads to an ``on-off" behavior
of conductance as discussed in Subsection \ref{subsec:sym}.
The second one is due to a collective effect of 
top-bottom and edge atomic interactions. As will be shown 
in Subsection \ref{subsec:valid},
the curvature at the edges induces large  $\pi-\sigma$ interactions, 
which may enhance or cancel the band-gap due to $\pi-\pi^*$ interactions 
between top and bottom atoms. This edge interaction, $\Delta\HH_{ED}$,
is defined such that the squashed CNT Hamiltonian is
$\HH_s=\HH_0+\Delta \HH_{TB} + \Delta\HH_{ED}$. In this regime $\Delta
\HH_{ED}$ induces $\pi-\sigma$ interaction and band gap cannot be
deduced within $\pi$-orbital model.   
  
Finally, in the third regime, where \dtb $<2.3$ \AA, we 
find that the band-gap
vanishes in all the simulations. 
At this stage the CNT undergoes
irreversible structural transformation where the honey-comb geometry
is lost. The electronic band structure of a completely squashed nanotube
resembles that of a metal, with multiple sub-bands passing through \EF.\\ 
Previous studies\cite{lu2003} predicted that the nanotube will always
become semiconducting in the course of  
squashing due to a spontaneous symmetry breaking.
However, we find that it is possible for a structure to remain mirror
symmetrical all the way till the irreversible regime as in the case of 
Fig.\ref{6-6-tube-gap}-a. Although mirror symmetry is broken
in the irreversible regime, this does not lead to the band-gap opening.

\subsection{The role of symmetry breaking in the 
formation of the band-gap \label{subsec:sym}} 

One can understand the band-gap opening in the first regime within a simple 
model\cite{lu2003}. In this model only a single \Pib-orbital/atom is
taken into account. In addition to this, squashing is treated as a
{\it first} order degenerate perturbation of the crossing
sub-bands, $\pi$ and $\pi^*$ of an ideal arm-chair CNT.
At the crossing point, the energy eigenstates are
determined by diagonalizing the perturbation Hamiltonian\cite{lu2003}:
\begin{equation}
\HH^p=
\left (
\begin{array}{ll}
V_{\pi\pi} & V_{\pi\pi^*}\\
V_{\pi^*\pi} & V_{\pi^*\pi^*}
\end{array}
\right ) \label{eq-perturbation}
\end{equation}
where $\HH^p=\HH_s-\HH_0$ and $\HH_s$,
$\HH_0$ are the Hamiltonians of squashed and ideal CNTs,
respectively. The diagonal term in Eq.\ref{eq-perturbation} results in
energy shift of $\pi$ and $\pi^*$ bands by $V_{\pi\pi}$ and
$V_{\pi^*\pi^*}$, respectively. This shifts the sub-bands crossing point
but does not lead to a band-gap. If the off-diagonal term is applied a band
gap \EG~opens at Fermi energy \EF, such that \EG$=2|V_{\pi\pi^*}|$.\\ 
In Fig.\ref{gap-total-all-pi}-a(b)  we show that this approximation
describes well the band-gap opening in the first regime (\dtb
$>3.8$ \AA) for results presented in Fig.\ref{6-6-tube-gap}-c(d).  
However, at slightly higher deformations
both the \Pib-orbital representation and the perturbation approach fail,
as elaborated in the next sub-section.\\
A band-gap opening in arm-chair nanotubes has been correlated to the
mirror symmetry breaking\cite{park1999,gulseren2002,lu2003}: when the
mirror symmetry is broken and bonds between top and bottom atoms are
formed, $V_{\pi\pi^*}$ becomes non {\it zero} and the band-gap opens up. 
The underlying mechanism is to make the two originally
equivalent sub-lattices \A~and \B~of the CNT distinguishable\cite{footnote1}. 
To investigate the rigor of the
correlation between the band-gap and 
mirror symmetry breaking, we have analyzed CNT atomic positions as a
function of squashing.  In each panel of Fig.\ref{6-6-tube-gap} we
display a snap-shot of {\it one} (6,6) CNT ring  at
\dtb~$=2.90\pm0.05$~ \AA. It is clear from Fig.\ref{6-6-tube-gap}-a,
that mirror symmetry is preserved and indeed the band-gap does not develop
in the course of squashing. However, mirror symmetry is broken for the
other simulations as indicated by the rest of the insets of
Fig.\ref{6-6-tube-gap}. Yet, our calculations, which 
include 4-orbitals/atom, show that
the band-gap vanishes at \dtb~$\sim 2.9$  \AA~for panels (b) and (c) of
Fig.\ref{6-6-tube-gap}. These results indicate clearly that
the degree to which sub-lattices
\A and \B are distinguishable is not the main factor in determining the
magnitude of the band-gap upon deformation.  We  conclude that {\it mirror 
symmetry breaking is necessary, but not sufficient for the band-gap 
opening}, and thus can't be used as a general guide to the 
metal-to-semiconductor transition. 

\subsection{Failure of \Pib-orbital representation 
and perturbation approach \label{subsec:valid}}

We now check the validity of the above model for the description of the 
band-gap. The $sp^3$ (4 orbitals/atom) representation of a CNT
Hamiltonian ($\HH_0$ or $\HH_s$) can be transformed
into {\it two} sub-blocks $\HH_{sp2}$ and $\HH_{\pi}$\cite{footnote2}. 
The band structure of an undeformed carbon nanotube in the
vicinity of the Fermi energy \EF, can be described using only 
the $\pi$-orbital Hamiltonian, $\HH_{\pi}$.  
The band-gap derived using the
single \Pib-orbital/atom model for these systems is shown in
Fig.\ref{gap-total-all-pi}-a,b as dashed lines. It is clear from
Fig.\ref{gap-total-all-pi}-a that, for \dtb~$<3.5$ \AA, the single
\Pib-orbital/atom model
overestimates the band-gap and does not predict the dip at
\dtb~$\sim2.9$  \AA. Similarly,  
it fails to determine band-gap \EG~for
\dtb~$<4.5$  \AA ~in Fig.\ref{gap-total-all-pi}-b.
These results show that the band structure is determined by
interactions between all orbitals, rather than $\pi$-orbitals alone.

We have also tested the rigor of perturbation theory, {\it i.e}, whether 
the band-gap  \EG~ can be estimated as $2|V_{\pi\pi^*}|$. 
In
Fig.\ref{gap-total-all-pi}, we display the value of 
$2|V_{\pi\pi^*}|$ calculated
using the full Hamiltonian (triangle down symbols) and 
the $\pi$-orbital Hamiltonian (triangle up). It is clear that
within $sp^3$ model perturbation theory results do not match \EG~ for
\dtb~$<3.5~(3.7)$ \AA~in Fig.\ref{gap-total-all-pi}-a(b). 
Even if a single \Pib-orbital/atom model is employed, 
perturbation theory results fail to describe \EG~ 
for \dtb~$<2.8~(3.0)$ \AA~in Fig.\ref{gap-total-all-pi}-a(b) . 
Hence we
conclude that 
neither the single \Pib-orbital/atom model nor perturbation 
theory, assumed in Eq.\ref{eq-perturbation} is able to 
describe the electronic band structure of a squashed arm-chair carbon nanotube.

To determine the origin of single \Pib-orbital/atom failure to
describe the electronic properties of squashed arm-chair CNT in the
vicinity of \EF, we decompose the perturbation Hamiltonian, $\HH_s-\HH_0$, 
to a sum of top-bottom and edge interaction, $\Delta\HH_{TB}$ and
$\Delta\HH_{ED}$.
We define
$\HH_{TB}=\HH_0+\Delta\HH_{TB}$ and use this Hamiltonian 
to find the band-gap, shown in Fig.\ref{gap-total-all-pi-2}-a(b),
for the structures discussed in Fig.\ref{gap-total-all-pi}-a(b).
We note that single \Pib-orbital/atom
model
describes \EG~due to all top-bottom interaction Hamiltonian accurately. 
Moreover,
this band-gap can be modeled within perturbation theory between $\pi$ and
$\pi^*$ states for \dtb $\ge 2.7$  \AA. If CNT is squashed such that
\dtb$\le 2.7$  \AA, top-bottom interaction is substantially enhanced and
perturbation theory fails to predict the band-gap. 

When similar analysis is
applied to the interactions at the edges, $\HH_{ED}=\HH_0+\Delta\HH_{ED}$, 
we find that
the 
single \Pib-orbital/atom model as well as perturbation theory fail to
reproduce full model calculations, as shown 
in Fig.\ref{gap-total-all-pi-2}-a'(b'). 
We see that  a  
single \Pib-orbital/atom model under-estimates 
the true band-gap \EG~ computed using 4 orbitals/atom, while 
the perturbation theory result $2|V_{\pi\pi^*}|$ predicts a 
much smaller value.
These graphs indicate that at the high 
curvature edge regions,
$\pi-\sigma$ interaction is large compared to $\pi-\pi^*$ interaction
and cannot be neglected.

The degree of sophistication required to describe 
the band-gap of a squashed CNT in the second regime can be summarized 
by the following steps: (i) if the mirror symmetry is broken, top bottom
interaction induces a band-gap opening, 
which can be predicted within the single 
\Pib-orbital/atom
model and a
perturbation theory; (ii) Further squashing increases 
$\pi-\sigma$ interaction at the edges due to large curvature. 
This interaction  necessitates the use of full calculation within $sp^3$ model.
It is also important to note that neither top-bottom nor 
edge interactions alone can explain the dips of the band-gap.
Only cancellation effect of both interactions leads 
to the oscillations of the band-gap in Fig.\ref{6-6-tube-gap}-b,c.

\subsection{ Effect of CNT diameter}

In this subsection we highlight the most important results for
arm-chair CNTs with larger diameters. In Fig.\ref{all-tubes-gap} 
we show the band-gap of $(n,n)$ nanotubes as a function of squashing,
where $n\in\{7,8,9,10,11,12\}$. 

In each panel of Fig.\ref{all-tubes-gap}, we display
a snap shot of the atomic configuration of the corresponding CNT. These
insets indicate that when mirror symmetry is strongly broken, larger
band-gap is formed. The origin of
this gap is of the same nature as in the (6,6) CNT. It is 
due to top-bottom interactions as well as edge effects. 
In Fig.\ref{all-tubes-gap}, we present
the gap results when including only interactions up-to {\it second}
nearest neighbor as continuous lines while dashed lines include all
Hamiltonian except top-bottom interactions. 
We note that the Hamiltonian which takes into account only {\it
first} and {\it second} nearest neighbors is similar to edge
interaction Hamiltonian $\HH_{ED}$.
These results indicate that edge
effect on the band-gap formation is relevant only for \dtb~$\le 3.5$ \AA.
The initiation of the gap at large separation distance, \dtb, is
solely due to top-bottom interactions as found for (6,6) CNTs.

Our simulations also show that {\it there is no preference for mirror
symmetry breaking during CNT squashing.} This is clearly seen for (7,7)
and (8,8) CNTs whose gap results are shown in
Fig.\ref{all-tubes-gap}-a,b, respectively. In the course of squashing
of these tubes, mirror symmetry is broken at \dtb~$\sim3.4$~\AA~and
a gap is formed around $E_F$. After $1-2$ steps mirror symmetry is
recovered and the gap closes at an early stage, as seen in the graphs of
Fig.\ref{all-tubes-gap}-a,b.

The major differences of the larger diameter nanotubes are
{\it (i)} the initiation of the band-gap at smaller deformation when
mirror symmetry is broken, and {\it (ii)} the increased 
stability of the system in the course of squashing.
This stability is due to robustness of the CNT conformation with
respect to the tip:
larger tubes have more atoms at the surface interacting with
graphene layers. Hence 
``tank-treading'' as well as translational motion is less significant.
Therefore, we expect that {\it for CNTs with large diameters,
the band-gap opening is more reproducible}.

\subsection{Effect of self consistency}

All our calculations are based on DF-TB
parameterization. However, such empirical potentials may suffer from
transferability problems, especially for largely deformed
structures. Hence we need to check the validity of our conclusions
using better models. To test our results, we consider the
calculations of (6,6) CNT structures displayed in
Fig.\ref{6-6-tube-gap}-a,b,d. 
Band-gap calculation using DF-TB model is compared to
calculations within self consistent density
functional tight
binding model (SCTB)\cite{elstner1998,frauenheim2000} and density
functional theory calculations (DFT)\cite{hohenberg1964}. The latter
is performed using Gaussian03
framework\cite{gaussian2003} within BPW91 exchange-correlation
functional parameterization\cite{burke1998} and 6-31G basis set.

In Fig.\ref{compare-TB-SC-DFT} we show the band-gap calculation results for
the three different simulations using different models.
When the CNT conserves its mirror symmetry a band-gap
cannot develop at $E_F$. 
The three models agree and the
results are identical as shown in
Fig.\ref{compare-TB-SC-DFT}-a. 
When  mirror symmetry is broken, the
three models predict a band-gap opening,
{\it non monotonic behavior} and closure
in the same region of deformation, but the magnitude
depends strongly on the
description of the interaction.
In particular, DF-TB results
which are shown as dotted lines with open circles in
Fig.\ref{compare-TB-SC-DFT} do not incorporate charge
redistribution. 
Such a description is reasonable for small
deformation. However as \dtb~decreases below $3.3$~ \AA, charge can
redistribute between carbon atoms and this may result in different
electronic properties. The importance of charge transfer revealed in
the SCTB calculations is shown as continuous
line in Fig.\ref{compare-TB-SC-DFT}. 
It diverges from DF-TB
model only under strong deformation, \dtb~$< 3.3$~ \AA.

The band-gap calculated within DFT is different
from DF-TB model. The origin of this difference is {\it two
fold}: DFT underestimates the band-gap and DF-TB model fails to describe long
range interaction. When mirror symmetry is broken, both DF-TB
and DFT predict mismatch between 
sub-lattices \A~and \B, but the strength of the mismatch depends on
the interaction.  Within DFT parameterization, the exchange 
correlation functional is underestimated and hence the band-gap is
smaller than DF-TB results under large deformations. 
However, at
\dtb~$\ge 3.5$~ \AA, DF-TB fails to describe long range interaction
due to an imposed cut-off radius of $5.2$ \AA.
This causes 
DFT results to be larger than those of DF-TB. Therefore, for large
diameter squashed CNTs with broken mirror symmetry a band-gap can
develop at much earlier stage, {\it i.e} at \dtb$>$ 5.0 \AA.

Finally we note that our results derived from the three models are in
qualitative agreement, and our conclusion deduced from DF-TB
calculations are valid. Quantitatively \EG~ depends on the model and
some differences emerge. In particular, within DFT model
conductance oscillations are less pronounced and disappear in
Fig.\ref{compare-TB-SC-DFT}-b. However, this model predicts \EG~
oscillations in Fig.\ref{compare-TB-SC-DFT}-c and the band-gap
vanishes \dtb$\sim$2.8 \AA.

\section{Conclusion \label{sec:conclu}}
We have performed the state of the art DF-TB MD simulations
of squashing arm-chair CNT. Our analyses have been carried out on
tubes with diameters in the range of $8-16$  \AA. 
Such a large number of simulations enable to have more conclusive
analysis on these systems and we find that:
\begin{itemize}
\item
Force required to squash CNT increases with tube diameter and is larger for
conformations with mirror symmetry preserved.
The path for squashing an arm-chair CNT can be split into {\it three}
different regimes. For \dtb$>3.8$ \AA, the force exerted on the AFM tip
is small ($<10$nN/per CNT unit cell) and the CNT undergoes most of the
compression. In the intermediate regime, \dtb$\sim 2.3-3.8$ \AA, force
is substantially enhanced and reaches $40-100$nN/per CNT unit
cell. Finally, for \dtb$<2.3$ \AA, CNT is under strong deformation. It
undergoes atomic relaxation and looses its hexagonal shape, resulting
in force drop-off.
\item
If CNT mirror symmetry is broken, \EG~can develop due to
mismatch between sub-lattices \A~and \B. However, this
distinguishability between the sub-lattices is due to top-bottom
interactions as well as edge effects.
The former can be modeled within single \Pib-orbital/atom while
the latter has strong $\pi-\sigma$ interaction and 
cannot be represented within single \Pib-orbital/atom.
\item
The band-gap is initiated primarily due to the top-bottom interactions, but for \dtb
$\le 3.5$ \AA~mismatch at the edges becomes important and can lead to
cancellation of the band-gap formed. This can result in band-gap
oscillation as a function of squashing. However this effect depends
on the exact atomic position at the edges. Hence we do not expect smooth
variation of the conductance as a function of squashing.
\item
Under strong deformation CNT looses its honey-comb structure and
becomes metallic.
\item
Large diameter CNTs have more contact area with AFM tip during
squashing. Hence, their conformation with respect to the tip is more
robust. Their squashing results are expected
to be more reproducible in an AFM experiment performed in the
reversible regime of the CNT structure.
\item
Band-gap formation of squashed arm-chair CNT described by DF-TB, SCTB and
DFT models are in qualitative agreement. However, quantitatively the value
of the band-gap is model dependent. Under large
deformation, \dtb~$<3.5$~ \AA, charge transfer is important and band-gap is
over estimated within DF-TB model. In addition to this, DF-TB model
under estimates long range interaction. Hence interaction between top
and bottom sides of CNT are under estimated and this results in smaller
gap compared to DFT for \dtb~$>3.5$~ \AA. Hence large diameter squashed
CNTs with broken mirror symmetry, are expected to develop a band-gap
even for \dtb~$\ge 5.0$ \AA. 
\end{itemize}
\section{Acknowledgment}
HM was supported by NASA contract to UARC/ELORET and NSERC of Canada.
AS and MPA was supported by NASA contract to UARC.
We are grateful to Dr. A. Maiti and Dr. J. Okeeffe
for useful discussions.

\newpage 
\begin{figure}
\caption{\label{distance-force} 
The inset in (b) displays the atomic configuration of a squashed CNT
between tip and sample. \dts, and \dtb~ are the tip-sample separation
distance and the distance between top and bottom sides of CNT,
respectively. (a) \dtb~ {\it vs.} \dts;
CNT squashing undergoes {\it three} regimes, \dtb$>3.8$ \AA,
$3.8$ \AA$>$\dtb$>2.3$ \AA, and \dtb$<2.3$ \AA. 
These regimes  are separated
by the horizontal dashed lines. In the first regime, the slope of \dtb~
{\it vs.} \dts~ is $\sim 0.88\pm
0.07$ while in the second regime the slope is $\sim 0.6\pm
0.05$, where the corresponding fits are represented as thick solid lines. 
At \dtb $<2.3$ \AA, CNT undergoes structural transformation and
the honey-comb structure cannot be identified. (b),(c) Force exerted on
the tip (top graphene sheet) as a function of \dts~and \dtb, respectively.
In all panels, every
thin line corresponds to one simulation and results for (6,6) and
(12,12) CNT are highlighted as thick dotted lines and dashed lines
respectively. 
}
\end{figure}

\begin{figure}
\caption{\label{6-6-tube-gap}\EG~ of (6,6) CNT at \EF~as a
function of \dtb~for different simulations. In (a) and (b) only top
graphene layer is displaced, while in (c) and (d) both layers are
moved to squash CNT. The insets in each panel show the atomic
configuration of the corresponding CNT unit cell for
\dtb~$2.90\pm0.05$ \AA~where grey atoms and dark grey atoms indicate
top and bottom sides of CNT, respectively. 
}
\end{figure}

\begin{figure}
\caption{\label{gap-total-all-pi}
(a) and (b) show band-gap at \EF~and $2|V_{\pi\pi^*}|$ of (6,6) CNT for
simulation results shown in Fig.\ref{6-6-tube-gap}-c,d, respectively. 
Continuous lines and dashed lines correspond to \EG~using 4
orbitals/atom and a single \Pib-orbital/atom, respectively. Thin lines 
with triangle down symbols and dashed lines with triangle-up symbols
represent $2|V_{\pi\pi^*}|$ using 4 orbitals/atom and a single
\Pib-orbital/atom, respectively.  
}
\end{figure}

\begin{figure}
\caption{\label{gap-total-all-pi-2}
(a,b) and (a',b') show band-gap at \EF~and $2|V_{\pi\pi^*}|$ of (6,6) CNT for
simulation results shown in Fig.\ref{6-6-tube-gap}-c,d due to
top-bottom interactions ($\HH_{TB}$) and edge effects ($\HH_{ED}$),
respectively. Top and bottom 
of CNT are shown in the insets of Fig.\ref{6-6-tube-gap} with 
grey and dark grey atoms. In these panels, continuous and dashed
lines correspond to \EG~using 4 orbitals/atom and a single
\Pib-orbital/atom, respectively. Thin lines with triangle down symbols
and dashed lines with triangle-up symbols represent $2|V_{\pi\pi^*}|$
using 4 orbitals/atom and a single \Pib-orbital/atom, respectively.
}
\end{figure}


\begin{figure}
\caption{\label{all-tubes-gap}Band-gap of arm-chair CNTs as a function
of squashing. (a), (b), (c),
(d), (e) and (f) correspond to (7,7), (8,8), (9,9), (10,10), (11,11)
and (12,12) tubes, respectively. Dotted lines with open circles
correspond to all interactions, continuous lines are interactions up-to the
{\it second} nearest neighbor and dashed lines correspond to results
when top-bottom interactions are omitted. 
The insets in each panel show the atomic configuration of
the corresponding CNT unit cell for \dtb~closest to $3.2$~ \AA~where grey
atoms and dark grey atoms indicate top and bottom sides of CNT,
respectively. 
}
\end{figure}

\begin{figure}
\caption{\label{compare-TB-SC-DFT}Energy gap of (6,6) CNT at $E_F$ as a
function of separation distance between top and bottom sides of CNT (\dtb)
for simulations shown in Fig.\ref{6-6-tube-gap}-a,b,d. Dotted lines with
open circles are the original calculations within DF-TB
model. Solid and dashed lines correspond to calculations using
SCTB and DFT, respectively.
}
\end{figure}


\begin{thebibliography}{200}
\bibitem{footnote1}
Two Carbon atoms define a graphene unit cell. Hence graphene can be viewed
as a superposition of two sub-lattices each composed of a single
atom. This defines sub-lattices \A~and \B~which have identical
interactions for ideal CNT.
\bibitem{footnote2}
To determine unit vector $\vec{U}$ corresponding to the
$\pi$-direction at every Carbon atom, we initially
find the unit vectors $\vec{C_1},\vec{C_2}$ and $\vec{C_3}$ in the
direction of the three nearest neighbors of the Carbon atom. Three
vectors, $\vec{V_1}, \vec{V_2}$ and $\vec{V_3}$ are defined such that
$\vec{V_i}=\epsilon_{ijk} (\vec{C_i}.\vec{C_j}) \vec{C_k}+ 
\epsilon_{ikj}(\vec{C_i}.\vec{C_k})\vec{C_j}$. $\vec{U}$ should be
orthogonal to the three vectors, $\vec{V_1}, \vec{V_2}$ and
$\vec{V_3}$; hence 
$\vec{U}$ is determined such that $[\vec{V_1}\,
\vec{V_2}\,\vec{V_3}].\vec{U}$ has negligible elements.
\end{thebibliography}
\end{document}